\documentclass[11pt]{article}
\usepackage{amsfonts}
\usepackage{amsmath, amssymb, amsthm}
\usepackage{fancyhdr}
\usepackage{graphicx}
\usepackage{dcolumn}
\usepackage{bm}

\begin{document}
{\setlength{\oddsidemargin}{1.4in}
\setlength{\evensidemargin}{1.4in} } \baselineskip 0.50cm
\begin{center}
{\LARGE {\bf Exact solutions of Einstein's equations \\ possessing dark energy}}
\end{center}

\begin{center}
Ng. Ibohal \\
Department of Mathematics, Manipur University,\\
Imphal - 795003, Manipur, India.\\
E-mail: ngibohal@iucaa.ernet.in
\end{center}

\begin{center}
Ngangbam Ishwarchandra\ddag \quad and \quad K. Yugindro Singh\dag\\
Department of Physics, Manipur University,\\
Imphal - 795003, Manipur, India\\
E-mail: \ddag \,ngishwarchandra@gmail.com, \dag\,yugindro361@gmail.com
\end{center}
\date{}

\vspace*{.25in}
\begin{abstract}
In this paper we propose a class of exact solutions (stationary and non-stationary) of Einstein's field equations. We find that the space-time geometries of the solutions are non-vacuum and conformally flat, whose  energy-momentum tensors possess dark energy  with negative pressure and the energy equation of state parameter $w=-1/2$. We also find that the time-like vector fields of the matter distributions of the solutions are expanding, shearing with acceleration and zero-twist. It is also found that, due to the negative pressure, the energy-momentum tensors violate the strong energy conditions leading to the repulsive gravitational fields of the space-time geometries. Energy-momentum tensors for the solutions also obey the energy conservation equations. From these physical properties of the matter distribution we may refer the space-times to as examples of exact solutions of the Einstein's field equations admitting dark energy  with negative pressure. It is to note that the approximate sizes of the masses of the (stationary and non-stationary) solutions are less than $(1/2)\times 10^{-60}$ in Bousso's length scale $r>10^{60}$. We also find that the surface gravities on the horizons are directly proportional to their respective masses. \\\\
{\bf Keywords:} Exact solutions; Einstein's equations; energy conditions; surface gravity; dark energy.

\end{abstract}
\vspace*{0.15in}
{\bf 1. Introduction}
\vspace*{0.15in}
\setcounter{equation}{0}
\renewcommand{\theequation}{1.\arabic{equation}}

Nowadays, it is found that the Wang-Wu mass function [1] plays a very important role in generating (embedded and non-embedded) exact solutions of Einstein's field equations. The mass function is expressed as a power series of the coordinate $r$ as
\begin{equation}
M(u,r)= \sum_{n=-\infty}^{+\infty} q_n(u)\,r^n,
\end{equation}
where $q_n(u)$ are arbitrary functions of retarded time coordinate $u$. The mass function is being utilized in generating {\sl non-rotating} embedded Vaidya solution  into other spaces by choosing the function $q_n(u)$ corresponding to the number $n$ [1]. Later on, the utilization of these mass functions has been extended in rotating system and found the role of the number $n$ in generating rotating embedded solutions of field equations [2]. The meaning of the power $n$ in the expansion series (1.1) for the known spherically axisymmetric solutions are as follows [1, 2]
\begin{itemize}
  \item[(i)] $n = 0$ corresponds to the term containing mass of the vacuum Kerr family solutions such as Schwarzschild, Kerr;
  \item[(ii)] $n = -1$ is equivalent to the charged term of Kerr family such as Reissner-Nordstrom, Kerr-Newman;
  \item[(iii)] $n = 1$ furnishes the term of the global monopole solution;
  \item[(iv)] $n = 3$ provides the de Sitter cosmological models, rotating and non-rotating.
\end{itemize}
These values of $n$ are conveniently used for stationary solutions (non-rotating and rotating). It is possible to obtain non-stationary Vaidya-Bonnor black holes, non-rotating [1] and rotating [2] when the values of $n = 0$ and $n =-1$ are summed up, since Vaidya-Bonnor metric describes a charged solution. When the above four values of $n$ $(= -1, 0, 1, 3)$ are together, one can find the charged Vaidya-de Sitter-monopole solution, non-rotating [1] and rotating [2]. Non-stationary rotating as well as non-rotating de Sitter cosmological models can also be obtained when $n=3$ [3]. In fact the Wang-Wu power series expansion of the mass function turns out to be the most convenient method to generate new viable (embedded or non-embedded) solutions [2] of the field equations if one uses Newman-Penrose (NP) formalism [4]. From the above identification of power $n$ we observe that the cases $n=2$ and $-2$ in the mass function (1.1) have not been seen considered before so far in the scenario of exact solutions of Einstein's field equations. This is the main aim of the paper for generating viable solutions of physical interest and to analyze the nature of the matter distributions in the space-time geometry. Here we shall concentrate only the case $n=2$. The other case $n=-2$ may be discussed elsewhere.

The important feature of the solutions with $n=2$ proposed here is that the space-time metrics are {\it non-asymptotically} flat when $r\rightarrow\infty$. However, they are flat at the origin $r\rightarrow 0$. Geometrically, it is acceptable that any curved space is locally flat. The idea of non-asymptotic flatness is in agreement with the Carter's suggestion [5,6] in generating exact solutions that ``it is not necessary to assume asymptotic flatness nor make the assumption that there are no other Killing vectors than $\xi=\partial_{t}$ and $\eta=\partial_{\phi}$'' [7]. Other examples of non-asymptotic space-times are (a) the de Sitter solution with cosmological constant and (b) rotating Kerr-NUT solution.

The paper is organized as follows: Sections 2 and 3 deal with the derivation of a class of viable solutions (stationary and non-stationary) of Einstein's equations. We find that the masses of the  solutions proposed here describe the gravitational fields of the space-time geometries and also determine the matter distributions with negative pressures, whose energy equations of state have the value $-1/2$. However, the energy-momentum tensors of the matter distributions with negative pressures do not describe a perfect fluid, which can be seen in the next sections. These non-perfect fluid distributions are in agreement with the remark of Islam [7] -- ``it is not necessarily true that the field is that of a star made of perfect fluid''. We also find that each solution has a coordinate singularity with  horizon. Consequently we discuss the areas, entropies and surface gravities at the horizons for the solutions. The existence of the horizons discussed here are also in accord with the cosmological horizon [8] of de Sitter space with constant $\Lambda$, which is usually considered to be a common  candidate of dark energy with the equation of state parameter $w = -1$. The paper is concluded in Section 4 with reasonable remarks and evolution of the solutions with the physical interpretation. Thus we summarize the results of the paper in the following theorems:

\newtheorem{theorem}{Theorem}
\begin{theorem}
An exact solution (stationary or non-stationary) admitting an energy-momentum tensor of a dark energy having negative pressure with equation of state parameter $w=-1/2$, is a non-vacuum, conformally flat {\rm (NCF)} space-time.
\end{theorem}
\begin{theorem}
The energy-momentum tensor of the matter distribution in an {\rm NCF} (stationary or non-stationary) space-time violates the strong energy condition leading to a repulsive gravitational force in the geometry.
\end{theorem}
\begin{theorem}
The time-like vector fields of the matter distributions in the (stationary and non-stationary) {\rm NCF} space-times are expanding, accelerating and shearing with zero-twist.
\end{theorem}
\begin{theorem}
The surface gravity at the horizon of an {\rm NCF} (stationary or non-stationary) space-time is directly proportional to the mass of the solution.
\end{theorem}
Theorem 1 shows the physical interpretation of the solutions that all  components of the Weyl tensors of the space-time metrics vanish indicating {\it conformally flatness} of the solutions. The energy-momentum tensors associated with the solutions admit dark energy having the negative pressures and the energy equation of state parameters $w=-1/2$. It is also found that the masses of the solutions not only describe the gravitational fields in the space-time geometries, but also measure the energy densities and the negative pressures in the energy-momentum tensors indicating the {\it non-vacuum} status of each solution. It is the assertion of General Relativity that ``the space-time geometry is influenced by the matter distribution'' [9]. Due to the negative pressure in the energy-momentum tensors, the violation of the strong energy condition is shown in Theorem 2 leading to a repulsive gravitational force in the space-time geometries. Theorem 3 shows the physical interpretation of time-like vector fields of the matter distributions. Theorem 4 indicates the existence of gravity on the horizons depending on their respective masses. It is to mention that Schwarzschild solution represents a vacuum, non-conformally flat space-time showing the difference from the non-vacuum, conformally flat (NCF) solutions discussed here. The presentation of the article is based on mathematical calculation for deriving exact solutions of Einstein's field equations. In this paper we utilize the differential form language developed by McIntosh and Hickman [10] in Newman-Penrose (NP) spin coefficient formalism [4] in $(-2)$ signature as mathematical tool.

\vspace*{.15in}

{\bf 2. An exact solution admitting non-perfect fluid}
\setcounter{equation}{0}
\renewcommand{\theequation}{2.\arabic{equation}}
\vspace*{.15in}

We consider a line element of a general canonical metric in Eddington-Finkestein coordinate systems $\{u, r, \theta, \phi\}$
\begin{eqnarray}
ds^2=\Big\{1-\frac{2}{r}M(u,r)\Big\} du^2+2du\,dr-r^2(d\theta^2+{\rm sin}^2\theta\,d\phi^2),
\end{eqnarray}
where $M(u,r)$ is referred to as the mass function and related to the gravitational fields within a given range of radius $r$. Here $u$ is the retarded time coordinate.

By virtue of the Einstein's field equations $R_{ab} - (1/2)Rg_{ab} = -KT_{ab}$ associated with the above line element (2.1), we find an energy-momentum tensor (stress-energy tensor) describing the matter distribution in the gravitational field as
\begin{eqnarray}
T_{ab}=\mu\ell_a\ell_b+2\,\rho\,\ell_{(a}\,n_{b)}
+2\,p\,m_{(a}\bar{m}_{b)},
\end{eqnarray}
where the quantities are found as
\begin{eqnarray}
\mu=-\frac{2}{Kr^2}M(u,r)_{,u}, \quad
\rho = \frac{2}{Kr^2}M(u,r)_{,r}, \quad p = -\,\frac{2}{Kr}M(u,r)_{,rr}
\end{eqnarray}
with the universal constant $K=8\pi G/c^4$. Here $\ell_a$, $n_a$ and $m_a$ are  given as follows
\begin{eqnarray}
&&\ell_a=\delta^1_a, \quad
n_a=\frac{1}{2}\Big\{1-2rM(u,r)\Big\}\,\delta^1_a+ \delta^2_a, \cr
&&m_a=-{r\over\surd 2}\,\Big\{\delta^3_a +i\,{\rm
sin}\,\theta\,\delta^4_a\Big\},
\end{eqnarray}
where $\ell_a$,\, $n_a$
are real null vectors and $m_a$ is complex having its conjugate $\bar {m}_a$ with the normalization conditions $\ell_an^a= 1 = -m_a\bar{m}^a$ and other inner products are zero. From (2.3) we observe that there is no straightforward way for solving the non-linear Einstein's field equations with the mass function $M(u,r)$ to generate a viable solution of physical interest. In order to have a meaningful physical interpretation of the solution (2.1) one has to consider some certain assumptions on the mass function $M(u,r)$ as the line element having the energy-momentum tensor (2.2) with the quantities (2.3) has no reasonable interpretation of exact solutions. For instance, the Vaidya null radiating solution can be obtained when one assumes the mass function to be $M(u,r)=M(u)$, leading to the condition $\rho=p=0$ in (2.3). Therefore, the mass function $M(u,r)$ for obtaining viable solutions can, without loss of generality, be expressed in the powers of $r$ as in (1.1) [1]. The above line element (2.1) with the mass function (1.1) includes most of the known solutions of Einstein's field equations, that can be seen with the identifications of power index $n=-1, 0, 1, 3$, depending on the system (rotating or non-rotating) mentioned in the introduction above. For example, de Sitter solution with cosmological constant $\Lambda$ is obtained by setting $q_{n}(u)=\Lambda/6$, when $n=3$ and $q_{n}(u)=0$, when $n\neq3$ in (1.1), providing the mass function $M(u,r)=(\Lambda/6)r^3$ and having energy density $\rho^* = \Lambda/K$, and pressure $p = -\Lambda/K$ in the non-rotating system [1]. Similarly, by assuming $q_{n}(u)= m$ when $n=0$, and $q_{n}(u)=0$ when $n\neq 0$, one can obtain the Schwarzschild solution with constant mass $m$ in the non-rotating coordinate system. On the other hand, this choice of the power $n=0$ provides the Kerr vacuum solution in a rotating system. This shows the fact that, although there is no straightforward way of solving the field equations with the mass function $M(u,r)$ in (2.1), the utilization of the Wang-Wu mass function (1.1) in the field equations, seems reasonable in generating viable solutions of physical interest. It is found that the case $n=2$ in the power series expansion (1.1) has not been considered before in the scenario of exact solutions of Einstein's field equations. It is hoped that this case $n=2$ may provide viable solutions of physical interest with reasonable interpretation of the matter distributions for both stationary as well as non-stationary space-times.

Here we consider the case $n=2$ in the mass function expansion (1.1) for generating a viable stationary solution of Einstein's field equations. Then we shall investigate the physical interpretation of the nature of the matter distribution in the space-time geometry. For this purpose we choose the Wang-Wu function $q_{n}(u)$ in $(1.1)$ as
\begin{eqnarray}
\begin{array}{cc}
q_n(u)=&\left\{\begin{array}{ll}
m, &{\rm when}\;\;n=2\\
0, &{\rm when }\;\;n\neq 2,
\end{array}\right.
\end{array}
\end{eqnarray}
such that the mass function takes the form
\begin{equation}
M(u,r)\equiv \sum_{n=-\infty}^{+\infty} q_n(u)\,r^n =m r^2,
\end{equation}
where $m$ is constant and $u =t-r$ is  the retarded time coordinate.
Using this mass function in (2.1)
we find a stationary line element
\begin{eqnarray}
ds^2=(1-2mr)du^2+2du\,dr-r^2(d\theta^2+{\rm sin}^2\theta\,d\phi^2),
\end{eqnarray}
where the constant $m$ is regarded as the mass of a test particle present in the space-time and is non-zero for the existence of the matter distribution in the  geometry. When $u$ = constant, the surface is the future directed null cone. This line-element describes a stationary solution, and has a coordinate singularity at $r=(2m)^{-1}$ describing a Lorentzian horizon. The line-element (2.7) is certainly different from (a) Schwarzschild solution with $g_{uu}=1-2M/r$ having singularity at $r=2M$, and $M$ being the Schwarzschild mass; and (b) de Sitter solution having a cosmological constant $\Lambda$ with $g_{uu}=1-(1/3)r^2\Lambda$ singularities at $r_{\pm}=\pm(3\Lambda)^{1/2}$.

Now using the mass function (2.6) in (2.2) and (2.3) we find an energy-momentum tensor (stress-energy tensor) describing the matter distribution in the gravitational field of the space-time geometry (2.7) as
\begin{eqnarray}
T_{ab} = 2\,\rho\,\ell_{(a}\,n_{b)}
+2\,p\,m_{(a}\bar{m}_{b)},
\end{eqnarray}
where the energy density $\rho$ and the pressure $p$ are found as
\begin{eqnarray}
&& \rho = \frac{4}{Kr}m, \quad p = -\frac{2}{Kr}m,
\end{eqnarray}
The null vector $n_a$ of (2.4) takes the form
\begin{eqnarray}
n_a=\frac{1}{2}\{1-2rm\}\,\delta^1_a+ \delta^2_a.
\end{eqnarray}
The equation (2.9) indicates that the contribution of the gravitational field  to the energy-momentum tensor $T_{ab}$ having the negative pressure $p$ is measured directly by the mass $m$ of a test particle.  Here from (2.7) and (2.9) we observe the key role of the mass $m$ that it not only describes the curvature of the space-time in (2.7) but also distributes the matter field (2.8) present in the space-time geometry.
Then we find the ratio of the pressure to the energy density as the equation of state for the solution
\begin{eqnarray}
\omega=\frac{p}{\rho}=-{1\over2}.
\end{eqnarray}
This negative value of $\omega$ is due to the negative pressure of the fluid.

The energy-momentum tensor (2.8) obeys the energy conservation laws, given in the Appendix A in terms of NP spin coefficients:
\begin{eqnarray}
T^{ab}_{\;\;\;\,;b}=0,
\end{eqnarray}
which shows the fact that the metric of the line element is a solution of Einstein's field equations. The components of energy-momentum tensor may be written as:
\begin{eqnarray}
T_{u}^{u}=T_{r}^{r}=\rho, \quad
T_{\theta}^{\theta}=T_{\phi}^{\phi}=-p
\end{eqnarray}
for future use.
We find the trace of the energy-momentum tensor
$T_{ab}$ (2.8) as follows
\begin{equation}
T=2(\rho-p)=\frac{12}{Kr}m.
\end{equation}
Here it is found that $\rho - p$ must always be greater than zero for the existence of the solution $(2.7)$ with $m \neq 0$, (if $\rho=p$ implies that $m$ will be vanished).
It is to emphasize that the energy-momentum tensor (2.8) does not describe a perfect fluid, i.e. for a perfect fluid, one has $T^{(\rm {pf})}_{ab} =
(\rho+p)u_au_b-p\,g_{ab}$ with a unit time-like vector $u_a$ and
its trace $T^{(\rm {pf})}=\rho-3p$, which is different from the one
given in (2.14).

{\bf Energy conditions:}
For the analysis of the energy conditions we shall introduce an orthonormal
tetrad with a unit time-like vector $u^a$ and three unit space-like vector fields
$v^a$, $w^a$, $z^a$ using the null tetrad (2.4) with (2.10) such as
\begin{subequations}
\label{eq:whole}
\begin{eqnarray}
&&u_a={1\over\surd 2}(\ell_a+n_a), \;\, v_a={1\over\surd
2}(\ell_a-n_a), \\ \label{subeq:1}
&&w_a={1\over\surd 2}(m_a+\overline m_a),\;\, z_a=-{i\over\surd
2}(m_a-\overline m_a), \label{subeq:2}
\end{eqnarray}
\end{subequations}
with the normalization conditions $u_au^a = 1$, $v_av^a$ =
$w_aw^a$ = $z_az^a = -1$ and other inner products being zero.
Then the metric tensor can be written as
\begin{equation}
g_{ab}=u_{a}u_{b}-v_{a}v_{b}-w_{a}w_{b}-z_{a}z_{b}.
\end{equation}
Now we consider a non-spacelike vector fields for an observer
\begin{eqnarray}
U_{a}=\hat{\alpha}u_{a}+\hat{\beta}v_{a}+\hat{\gamma}w_{a}+\hat{\delta}z_{a},
\end{eqnarray}
where $\hat{\alpha}$, $\hat{\beta}$, $\hat{\gamma}$ and $\hat{\delta}$ are arbitrary constants [11], subjected to the condition that
\begin{eqnarray}
U^{a}U_{a}=\hat{\alpha}^2-\hat{\beta}^2-\hat{\gamma}^2-
\hat{\delta}^2\geq0.
\end{eqnarray}
Then the energy-momentum tensor (2.8) can be written in terms of the orthonormal tetrad vectors given in (2.15) as
\begin{eqnarray}
T_{ab}=(\rho+p)(u_{a}u_{b}-v_{a}v_{b})-pg_{ab}.
\end{eqnarray}
Now $T_{ab}\,U^aU^b$ will represent the energy density as measured by the observer with the tangent vector $U^a$ (2.17). Then we have the following energy conditions:
\begin{itemize}
 \item[(a)] {\it Weak energy condition}: The energy-momentum   tensor  obeys  the inequality\,\,\,\,  $T_{ab}U^aU^b \geq 0$ for any time-like vector $U^a$ which
     implies that
     \begin{eqnarray}
     \rho>0, \quad \rho+p>0.
     \end{eqnarray}
 \item[(b)] {\it Strong energy condition}: The Ricci tensor for $T_{ab}$ satisfies the inequality \\ $R_{ab}\,U^aU^b\geq0$ for all time-like vector $U^a$, i.e. $T_{ab}U^aU^b\geq(1/2)T$, which yields
     \begin{eqnarray}
     p>0,\quad \rho+p>0.
     \end{eqnarray}
 \item[(c)] {\it Dominant energy condition}: For all future directed vector
     $U^a, T_{ab}U^b$ should be a future directed non-space like vector.
This condition is equivalent to
     \begin{eqnarray}
     {\rho}^2>0, \quad  {\rho}^2-p^2>0.
     \end{eqnarray}
\end{itemize}
Here, we find that the value of $p$ given in (2.9) does not satisfy the strong energy condition which has to be (2.21) as per the condition $T_{ab}U^aU^b\geq(1/2)T$ with (2.19). This violation of the strong energy condition is due to the negative pressure, and may lead to a repulsive gravitational force of the matter field in the space-time. It is like the repulsive cosmological constant [8] to lead to the accelerated expansion of de Sitter space. This proves the stationary part of Theorem 2 above.

We also find that the line element (2.7) of the space-time is {\it conformally flat} $C^{a}_{\;\;bcd}=0$,  i.e. all the tetrad components of Weyl tensor are  vanished
\begin{eqnarray}
\psi_0 =\psi_1 =\psi_2 =\psi_3 =\psi_4 =0.
\end{eqnarray}
 The curvature invariant for the solution (2.7) is found as
\begin{eqnarray}
R_{abcd}R^{abcd}=-\,\frac{160}{r^2}\,m^2,
\end{eqnarray}
which is regular on the `singular' surface $r=(2m)^{-1}$. This invariant diverges only at the origin. This indicates that the origin $r=0$ is a physical singularity. This shows that the singularity of the solution (2.7) at $r=(2m)^{-1}$ is caused due to the coordinate system, just like in Schwarzschild solution with the horizon $r = 2m$ [12]. From (2.7), (2.9) and (2.23), we come to the conclusion of the proof of Theorem 1 stated in the case of stationary part above.

\vspace*{.15in}
{\bf Raychaudhuri equation}: Here we shall analyze the nature of the time-like vector fields $u^a$ appeared in the energy-momentum tensor (2.19) for the stationary solution (2.7). The time-like vector $u^a$ is often considered to be the 4-velocity of a fluid. So the 4-velocity vector measures the kinematical properties of a fluid whether the fluid flow is expanding ($\Theta=u^a_{\:\,;a} \neq 0$), accelerating ($\dot{u}_a=u_{a;b}u^b\neq 0$), shearing $\sigma_{ab}\neq0$ or non-rotating ($w_{ab}=0$). We shall investigate the change of the volume expansion from the Raychaudhuri equation, such that we can understand how the negative pressure of the fluid affects the expansion of the solution. For this purpose we find the covariant derivative of the 4-velocity vector $u^a$ in terms of null tetrad vectors (2.4) with (2.10) as follows
\begin{equation}
u_{a;b}=\frac{1}{\surd 2}\Big\{m(\ell_a\ell_b-n_a\ell_b)-\frac{1}{r}(1+2mr)m_{(a}\bar{m}_{b)}\Big\},
\end{equation}
where $m$ is the mass of the solution. In deriving the above expression (2.25) we use the definition of $u^a$ given in (2.15a). This expression of $u_{a;b}$ is convenient to calculate the (volume) expansion scalar $\Theta=u^a_{\:\,;a}$ and acceleration vector $\dot{u}_a=u_{a;b}u^b$ as follows
\begin{subequations}
\label{eq:whole}
\begin{eqnarray}
&&\Theta\equiv u^a_{\:\,;a} = \frac{1}{\surd 2\,r}(1+3rm) \\ \label{subeq:1}
&&\dot{u}_a=-\frac{1}{2}m(\ell_a - n_a) \\ \label{subeq:2}
&&\dot{u}^a_{\:\,;a}=-\frac{3m}{2r}(1-mr). \label{subeq:3}
\end{eqnarray}
\end{subequations}
We find that the vorticity tensor $w_{ab}=u_{[a;b]}-\dot{u}_{[a}u_{b]}$ is vanished for the solution (2.7), and however, the shear tensor $\sigma_{ab}=u_{(a;b)}-\dot{u}_{(a}u_{b)}-(1/3)\Theta h_{ab}$ exists as
\begin{equation}
\sigma_{ab}=\frac{1}{6\surd 2 r}\Big[(\ell_a\ell_b+n_a n_b)-2\{\ell_{(a}n_{b)}+m_{(a}\bar{m}_{b)}\}\Big],
\end{equation}
which is orthogonal to $u^a$ ({\it i.e.,} $\sigma_{ab}u^b=0$). It is found that the mass $m$ of the solution does not explicitly involve in the expression  of $\sigma_{ab}$ but its involvement can be seen in the null vector $n_a$ given in (2.4). However, the mass $m$ directly determines the expansion $\Theta$ as well as the acceleration $\dot{u}_a$ as seen in (2.26). We find from (2.26) that the particle moving on the space-time geometry (2.7) follows the non-geodesic path of the time-like vector ($u_{a:b}u^b\neq 0$). This establishes the key role of the expansion of the solution with acceleration. The vanishing of the vorticity tensor $w_{ab}=0$ may be interpreted physically  as saying that the matter field of the solution is  {\it twist-free} (non-rotating) as mentioned earlier. This comes to the conclusion of the proof of Theorem 3 for the stationary case.

Now let us observe the consequence of the Raychaudhuri equation for the stationary solution (2.7). The Raychaudhuri equation is given by
\begin{equation}
\dot{\Theta}=\dot{u}^a_{\:\,;a}+2(w^2-\sigma^2)-\frac{1}{3}\Theta^2 +R_{ab}u^a u^b
\end{equation}
where $\dot{\Theta}=\Theta_{;a}u^{a}$. The shear and vorticity magnitudes are $2\sigma^2=\sigma_{ab}\sigma^{ab}$ and $2w^2=w_{ab}w^{ab}$; $R_{ab}$ is the Ricci tensor associated with the space-time metric $g_{ab}$. Then the Raychaudhuri equation for the  twist-free time-like vector is found as follows
\begin{equation}
\dot{\Theta}=-\frac{1}{4r^2}(1+2rm),
\end{equation}
which takes the constant value $\dot{\Theta}=-2m^2$ on the horizon $r=(2m)^{-1}$,
showing the constancy of the expansion along the time-like vector $u^a$. The impact of the negative pressure $p$ given in (2.9) is taken care in Ricci tensor $R_{ab}$ in (2.28). The acceleration vector  $\dot{u}_a$ and its scalar $\dot{u}^a_{\:\,;a}$ are negative, affecting the expansion rate in (2.28). It is to mention that the time-like vector $u^a$ is {\it not} a static Killing vector $\pounds_{u}g_{ab}\neq 0$, indicating the difference from any static time-like Killing vector $\xi^a$, ($\pounds_{\xi}g_{ab}=0$) which has no expansion and shear.

\vspace*{.23in}
{\bf Surface Gravity:}
The line element of the stationary solution (2.7) has a horizon at $r=(2m)^{-1}$. In this regard we have to note the fact that the existence of the horizon is in accord with the cosmological horizon of de Sitter space with constant $\Lambda$ [8,13], since it is regarded as a  common example of dark energy with the equation of state parameter $w = -1$. Therefore, we expect that it is highly important to observe the interpretation of the mass of the solution in connection with the area, entropy, surface gravity as well as the temperature for the horizon. So we find the area at the horizon $r=(2m)^{-1}$ as
\begin{eqnarray}
A=\int_0^\pi\int_0^{2\pi}(g_{\theta\theta}\,g_{\phi\phi})^{\frac{1}{2}}\,d\theta\,
d\phi\Big|_{r=(2m)^{-1}}=\pi m^{-2},
\end{eqnarray}
and the entropy, from the entropy-area relation $S=A/4$ [8], as
\begin{eqnarray}
S=\frac{1}{4}\pi m^{-2}.
\end{eqnarray}
It indicates that the area and entropy will always exist for the solution with non-zero mass $m$.

According to Carter [14] and York [15], the surface gravity ${\cal \kappa}$ of a horizon is defined by the relation $n^b\nabla_b n^a={\cal \kappa}n^a$, where the null vector $n^a$ in (2.10) above is parameterized by the coordinate $u$, such that $d/du=n^b\nabla_b$. Then the surface gravity is expressed in terms of NP spin-coefficient $\gamma$ as follows [16]
\begin{eqnarray}
\kappa=n^b\nabla_b\,n^a\ell_a=-(\gamma+\bar{\gamma}),
\end{eqnarray}
where $\gamma=-{m/2}$.
From this we find the surface gravity on the horizon $r=(2m)^{-1}$ as
\begin{equation}
\kappa = m,
\end{equation}
which shows that the surface gravity is directly measured by the mass, or in other words, it is directly proportional to the mass of the solution. This establishes the proof of Theorem 4 in the case of stationary solution. Then the Bekeinstein-Hawking temperature for the model at the horizon is found as
\begin{equation}
T=\frac{\hbar\kappa}{2\pi Gkc}=\frac{\hbar m}{2\pi Gkc},
\end{equation}
where $\hbar$ is the reduced Planck constant, $c$ the speed of light, $k$ the Boltzmann constant, and G the gravitational constant.
It indicates that the surface gravity and temperature of the horizon will never become zero for the existence of the stationary NCF solution $(m\neq0)$. When the mass $m=0$ becomes zero, the line element (2.7) will be a flat metric, the non-existence of the solution, and at this stage the surface gravity as well as the temperature will vanish. It is consistent with the  property of flat space-time geometry, where there is no gravity, one cannot determine the surface gravity of the solution. It is noted that the surface gravity $\kappa_{\rm Sch}$ of the Schwarzschild black hole is inversely proportional to the mass $M$ as $\kappa_{\rm Sch}=(4M)^{-1}$ on its horizon $r=2M$ [17].

\vspace*{.15in}
{\bf Size of the mass:} It is quite interesting to introduce a possible size of the mass of the solution discussed here. According to Bousso [13], stars are as distance as billions of light years, so $r>10^{60}$ and stars are as old as billions of years, $t>10^{60}$. In this length scale, the size of the mass of the solution at the horizon $r=(2m)^{-1}$ may become
\begin{eqnarray}
m=\frac{1}{2}r^{-1}<\frac{1}{2}\times10^{-60},
\end{eqnarray}
which is slightly bigger than the size of the cosmological constant $|\Lambda|\leq3r_{\Lambda}^{-2}\leq 3\times10^{-120}$ with the cosmological horizon $r_\Lambda=\sqrt{3/\Lambda}$ [13]. The relation (2.35) may provide an example of a tiny test particle having a small size mass in the Universe.

\vspace*{.15in}
{\bf Kerr-Schild ansatz}:
The line element (2.7) can be expressed in Kerr-Schild ansatz on Minkowski flat background $\eta_{ab}$ as
\begin{eqnarray}
ds^2&=&dt^2-dr^2-r^2(d\theta^2+{\rm sin}^2\theta\,d\phi^2)-2mrdu^2
\end{eqnarray}
under the transformation $t=u+r$. This is the Kerr-Schild ansatz on the flat background $\eta_{ab}$ in spherical coordinate system
\begin{eqnarray}
g_{ab}=\eta_{ab}+2Q\ell_{a}\ell_{b},
\end{eqnarray}
where $Q=-mr$, $g_{ab}$ is the metric tensor of the solution and  $\ell_{a}=\delta^u_a$ is the null vector with respect to $g_{ab}$ and $\eta_{ab}$ given in (2.4). This confirms the fact that the stationary solution obtained here is an exact solution of Einstein's field equations. The line element (2.7) can also be written in the $(t,r,\theta,\phi)$ coordinate system for future use as follows
\begin{eqnarray}
ds^2=(1-2mr)dt^2-(1-2mr)^{-1}dr^2-r^2d\theta^2-r^2{\rm sin}^2\theta\,d\phi^2,
\end{eqnarray}
under the transformation $dt=du+\{1/(1-2mr)\}dr$.
This is a very familiar form of the line element in General Relativity, and its determinant is $|g|=-r^4\sin^2\theta$. In this coordinate system we can easily observe the singularity at the point $r=(2m)^{-1}$.

\vspace*{.15in}
{\bf Physical interpretation of the matter distribution}: In order to interpret the physical meaning of the negative pressure of the matter distribution in the NCF solution, we shall consider other space-times having negative pressure and the energy equation of state with minus sign. For instance, the cosmological constant $\Lambda$ of the non-rotating de Sitter solution, whose energy-momentum tensor is
$T_{ab}^{\rm dS}=\Lambda g_{ab}^{\rm dS}$, is regarded as a common candidate
of dark energy [13,19-27] having the negative pressure $p=-\Lambda/K$, and the energy density $\rho=\Lambda/K$ with the equation of state parameter $w=p/\rho=-1$, where $K=8\pi\,G/c^4$.  $T_{ab}^{\rm dS}$ violates the strong energy condition leading to the term -- the {\it repulsive} (not attractive) cosmological constant $\Lambda$ [8]. This equation of state $w=-1$ is also satisfied for (a) the non-rotating non-stationary (time dependent) de Sitter with cosmological function $\Lambda(u)$, whose $T_{ab}=-\frac{1}{3}r\Lambda(u)_{,u}\ell_a\ell_b +
\Lambda(u)g_{ab}$ with $p=-\Lambda(u)/K$, $\rho=\Lambda(u)/K$  and $u$ is the retarded time coordinate [3]; (b) the rotating de Sitter solution $\Lambda(u)$ [3] having $p=-r^2\Lambda(u)(r^2+2a^2\cos^2\theta)/(KR^2R^2)$ and $\rho=r^4\Lambda(u)/(KR^2R^2)$ at the poles $\theta=\pi/2$ or $3\pi/2$, where $R^2=r^2+a^2\cos^2\theta$ with the non-zero rotational parameter $a$. It includes the case of the rotating stationary (time independent) de Sitter solution with constant $\Lambda$ [2]. This indicates that in the study of dark energy problems one needs not concentrate only on the cosmological constant $\Lambda$, and that one can consider the {\it non-constant} cosmological function $\Lambda(u)$ in the rotating as well as non-rotating de Sitter space-time geometries with the equation of state parameter $w=-1$ [2,3]. It is to mention the equation of state $w$ for ordinary matter field distributions for better understanding the dark energy problem. That we have the equation of state for other ordinary matters having positive sign (i) $w = 1$ for electromagnetic field having $\rho = p = e^2/(KR^2R^2)$ of the Kerr-Newman black hole with the constant electrical charge $e$, and in Vaidya-Bonner radiating black hole $\rho=p=e(u)^2/(KR^2R^2)$ with variable charge $e(u)$ of retarded time coordinate $u$ [2], (ii) $w=1/3$ for radiation field with $\rho-3p=0$ [13,19-27]. This shows the fact that dark energy always has a minus sign in the value of the energy equation of state parameter; whereas the ordinary matter has a positive sign. The negative sign in the equation of state is an important property for any matter field distribution to be interpreted as a {\it dark energy}.

The observations of luminosity-redshift relation for the type Ia supernovas [28-30] suggest that the missing energy should possess negative pressure $p$ and the equation of state $w=p/\rho$ [31]. The negative pressure of the dark energy may be the cause of the acceleration of the present Universe. Although the dark energy has been sought in a wide range of physical phenomena depending on the value of the equation of state parameter $w$, (i) a quintessence field $-1<w<-1/3$, (ii) the cosmological constant $w=-1$, (iii) a phantom field $w<-1$ [32,33], the nature of the dark energy still remains a complete mystery  [33,34] without any proper space-time geometry, except the assumption of a line element of a perfect homogeneous and isotropic space-time having a compatible energy-momentum tensor of perfect fluid $T^a_{\;\,b}={\rm diag}\{\rho(t), - p(t), - p(t), - p(t)\}$,
with $\rho(t)$ and $p(t)$ being the energy density and pressure of the matter distribution in the Friedmann-Robertson-Walker universe.

From the above scenario of dark energy and with the findings here (i) the negative pressure (2.9), (ii) the energy equation of state with minus sign $w=-1/2$ (2.11), (iii) the violation of strong energy condition (2.21), and (iv) the accelerating expansion of the time-like vector (2.26) for the stationary NCF solution, we may regard the matter distribution (2.19) as an example of {\it dark energy} with negative pressure whose space-time geometry is the line element (2.7).

\vspace*{0.15in}
{\bf 3. Non-stationary solution with negative pressure}
\setcounter{equation}{0}
\renewcommand{\theequation}{3.\arabic{equation}}
\vspace*{.15in}

In this section we shall develop a non-stationary (time dependent) version of the non-vacuum, comformally flat (NCF) space-time (2.7) discussed above. Since there is no straightforward way for solving the non-linear Einstein's field equations associated with the mass function $M(u,r)$ in (2.1) with (2.3) for a viable solution of non-stationary NCF space-time, we follow the Wang-Wu technique [1] as above. Therefore, we assume the Wang-Wu function having a variable mass $m(u)$ as follows:
\begin{equation}
M(u,r)\equiv \sum_{n=-\infty}^{+\infty} q_n(u)\,r^n =m(u)\,r^2
\end{equation}
when $n=2$.  Utilizing this mass function in the general canonical metric (2.1) we find a non-stationary line element as
\begin{eqnarray}
ds^2=\{1-2r\,m(u)\}\,du^2 +2du\,dr -r^2(d\theta^2+{\rm
sin}^2\theta\,d\phi^2).
\end{eqnarray}
where $m(u)$ is considered to be a variable mass of a test particle in the non-stationary system. The above solution has a coordinate singularity at $r=\{2m(u)\}^{-1}$. Such a generation of a non-stationary solution from the stationary one (2.7) is acceptable in the framework of General Relativity that the non-stationary Vaidya null radiating solution with variable mass $M(u)$ is a generalization of the stationary Schwarzschild vacuum solution with constant mass $M$.
Similarly, the non-stationary de Sitter solution [3] with a cosmological function $\Lambda(u)$ having horizons at $r_{\pm}=\pm\{3\Lambda(u)\}^{1/2}$ can be obtained from the stationary de Sitter model of constant $\Lambda$.

Now using the mass function (3.1) in (2.3) we find the null density $\mu$, the energy density $\rho$ and the pressure $p$ as
\begin{eqnarray}
\mu=-\frac{2}{K}m(u)_{,u}, \quad
\rho = {4\over Kr}m(u), \quad p = -{2\over Kr}m(u),
\end{eqnarray}
associated with a energy-momentum tensor for a non-stationary matter distribution:
\begin{eqnarray}
T_{ab}=\mu\ell_a\ell_b+(\rho+p)(u_{a}u_{b}-v_{a}v_{b})-pg_{ab}.
\end{eqnarray}
Here $\ell_a$ is the real null vector, $u_a$ a unit time-like vector $u_a u^a=1$ and $v_a$ a unit space-like vector $v_a v^a =-1$ defined as in (2.15a). It is observed that the presence of the null density $\mu$ in $T_{ab}$ is due to the non-constant mass $m(u)$ in the field equations showing the evolution of a non-stationary solution. From (3.3) it follows that the energy equation of state parameter $w=p/\rho$ has the value $-1/2$ with minus sign. If one sets the mass function $m(u)$ to a constant $m$, the above energy-momentum tensor will become the one given in (2.19) for the stationary solution with the quantities $\rho$ and $p$. The energy-momentum tensor obeys the energy conservation laws $T^{ab}_{\;\;\;\,;b}=0$ as shown in Appendix below. It shows the fact that the non-stationary solution (3.2) is an exact solution of Einstein's field equations.
It is also found that, due to the negative pressure, the $T_{ab}$ violates the {\it strong-energy condition},
\begin{equation}
\frac{1}{2}\mu+p\geq 0, \quad \frac{1}{2}\mu+\rho+p\geq 0.
\end{equation}
This violation indicates that the gravitational force of the non-stationary model is repulsive as in the case of stationary one (2.7) above which completes the proof of Theorem 2. However, $T_{ab}$ (3.4) with the negative pressure satisfies the {\it week} energy condition (i) ${\mu/2} + \rho \geq 0$, (ii) ${\mu/2} +
\rho +p \geq 0$, and {\it dominant} energy condition (i) $\mu\rho+\rho^{2}\geq 0$,
(ii) $-\mu\rho+\rho^{2}-p^2\geq 0$. Here we observe the difference between the strong energy conditions of the stationary (2.21) and that of the non-stationary (3.5) with the null density $\mu$. We also find that the non-stationary solution is {\it conformally} flat $C^a_{\;bcd}=0$, {\it i.e.}
\begin{equation}
\psi_0 =\psi_1 =\psi_2 =\psi_3 =\psi_4 =0.
\end{equation}
This shows that the solution  (3.2) with a mass function $m(u)$ has the same characterization of conformally flatness of stationary solution (2.21) with constant mass. Equations (3.3), (3.4) and (3.6) provide the proof of the non-stationary part of Theorem 1. It is also noted that the structure equation of the Riemann curvature invariant for variable mass (3.2) has a similar form of that of the solution (2.7) in (2.24) as
\begin{eqnarray}
R_{abcd}R^{abcd}=-\,\frac{160}{r^2}\,m^2(u),
\end{eqnarray}
which is regular on the `singular' surface $r=\{2m(u)\}^{-1}$. This invariant is divergent only at the origin $r=0$, which is a physical singularity.

We also have the expansion scalar $\Theta$ and the acceleration vector $\dot{u}_a$ with its scalar $\dot{u}^a_{\:\,;a}$ for the time-like vector $u_a$ appeared in (3.4)
\begin{subequations}
\label{eq:whole}
\begin{eqnarray}
&&\Theta\equiv u^a_{\:\,;a} = \frac{1}{\surd 2\,r}\{1+3rm(u)\} \\ \label{subeq:1}
&&\dot{u}_a=-\frac{1}{2}m(u)\{\ell_a - n_a\} \\ \label{subeq:2}
&&\dot{u}^a_{\:\,;a}=\frac{1}{2}m(u)_{,u}-\frac{3m(u)}{2r}\{1-m(u)r\}. \label{subeq:3}
\end{eqnarray}
\end{subequations}
However the shear tensor $\sigma_{ab}$ remains unchanged as in the stationary solution (2.24) and  the rotation tensor $w_{ab}$ is vanished (zero-twist). This follows the proof of Theorem 2 of the non-stationary case. From the Raychoudhuri equation, we have the rate of change of the expansion scalar as follows
\begin{equation}
\dot{\Theta}=\frac{1}{2}m(u)_{,u}-\frac{1}{4r^2}\{1+2rm(u)\}.
\end{equation}
The surface gravity of the non-stationary solution at the horizon $r=\{2m(u)\}^{-1}$ takes the form
\begin{equation}
\kappa = m(u)
\end{equation}
which shows the proof of Theorem 4 of non-stationary part. We have also seen the evolution of the non-stationary solution with the mass function $m(u)$ in (3.3), (3.8c) and (3.9). From (3.3), (3.5) and (3.8) we may regard the solution (3.2) as an example of a non-stationery space-time admitting an energy-momentum tensor of a dark energy with negative pressure having an equation of state parameter $w=p/\rho=-1/2$.

\vspace*{.15in}
{\bf 4. Conclusion}
\vspace*{.15in}

In this paper we develop a class of exact (stationary and non-stationary) solutions  of Einstein's field equations describing non-vacuum and conformally flat space-times, whose energy-momentum tensors possess dark energy fluids with the negative pressure and the equation of state parameter $w=-1/2$. The most exotic property of these solutions is that the  metrics describe both the background space-time structure and the dynamical aspects of the gravitational field in the form of the energy-momentum tensors. That is to say that the masses of the solutions play the role of both the curvature of the space-time ({\it non-flat}) as well as the source of the energy-momentum tensor with $T_{ab}\neq 0$ ({\it non-vacuum}) measuring the energy density and the negative pressure. This indicates that when we set the masses of the solutions to be zero, the space-times will become the flat Minkowski space with vacuum structure $T_{ab}=0$. In the case of Schwarzschild solution, the mass plays only the role of curvature of the space-time and cannot determine the energy-momentum tensor. That is why the Schwarzschild solution is a curved {\it non-flat, vacuum} space-time with $T_{ab}=0$. Here lies the advantage of the solutions (2.7) and (3.2) as {\it non-flat} and {\it non-vacuum} space-times over the Schwarzschild. It is also to mention that the solutions discussed here provide examples of conformally flat space-times, while other examples of conformally flat solutions are the non-rotating de Sitter models with cosmological constant $\Lambda$ [18], function $\Lambda(u)$ [3] and the Robertson-Walker metric [12].

We find that the time-like vector of the source is expanding $\Theta \neq 0$, accelerating $\dot{u}_a\neq 0$ (2.26b) as well as shearing $\sigma_{ab} \neq 0$ (2.27), but non-rotating $w_{ab}=0$. This means that the stationary observer of the solution does not follow the time-like geodesic path as $u_{a;b}u^b\neq 0$. Similarly, a non-stationary observer in (3.2) follows the non-geodesic path (3.8b). We also find that the energy-momentum tensors for the solutions violate the strong energy conditions. The violation of strong energy conditions is due to the negative pressure of the matter fields content in the space-time geometries, which can be seen in (2.21) and (3.5) above, and is not an assumption to obtain the solution (like other models mentioned in [33]). This violation indicates that the gravitational fields of the solutions are repulsive (as pointed out in [35]) leading to the accelerated expansions of the universe. The expansion of the space-time with acceleration is in agreement with the observational data [28-30]. It is also noted that the strong energy conditions for the stationary solution (2.21) associated with the stress-energy momentum tensor (2.19) and that of the non-stationary one (3.3) are different from that of the perfect fluid ($\rho \geq 0$, $\rho + 3p \geq 0$). This indicates that the strong energy condition is mainly depended upon the structure equation of a particular energy-momentum tensor.

It is emphasized the fact that our approach in the development of the solutions here is necessarily based on the identification of the power $n=2$ in the Wang-Wu mass function without any extra assumption. This identification of the power $n=2$ in the mass function (1.1) has considered here for the first time, and not been seen discussed before in the scenario of exact solutions of Einstein's field equations. It is also noted that we do not consider the Friedman-Robertson-Walker metric, filled with perfect fluid, which is assumed to be the standard approach for the investigation of dark energy problem as mentioned earlier [13,19-26]. That is why the energy-momentum tensors associated with the solution (2.7) and (3.2) do not describe a perfect fluid. This fact can be observed from the trace $T=2(\rho-p)$ of the $T_{ab}$ for both stationary and non-stationary solutions. This non-perfect fluid  distribution of the solutions is also in accord with Islam's suggestion that it is not necessarily true that stars are made of perfect fluid [7].

From the study of the above solutions, we find that the energy densities are only contributed  from the masses of the matters. It is the fact that without the mass of the solutions, one cannot measure the energy density and the negative pressure of the energy-momentum tensors in order to obtain the energy equation of state $w=-1/2$. This means that the negative pressures and the energy densities associated with the energy-momentum tensors (2.19) and (3.4) are measured by the masses that produce the gravitational field in the space-time geometries of the solutions. Hence, we may conclude that the stationary and non-stationary solutions may explain the essential part of Mach's principle -- ``The matter distribution influences the space-time geometry'' [36]. It is emphasized that the equations of state parameters $w=-1/2$ for the matter distributions (2.19) and (3.4) are belonged to the range $-1<w<0$ focussed for the best fit with cosmological observations in [25] and references there in.

The metrics appear singular when $r=0$ and $r=(2m)^{-1}$ for stationary and $r=\{2m(u)\}^{-1}$ for non-stationary at a particular value of $u$. These values of $r$ have special importance. At the origin $r=0$, there is a physical singularity  where the curvature invariants diverge as shown in (2.24) and (3.7); at $r=(2m)^{-1}$ for stationary and $r=\{2m(u)\}^{-1}$ for non-stationary, the invariants are well behaved and finite. Accordingly, we find areas, entropies as well as surface gravities at the horizons. It is found that the surface gravities given in (2.33) and (3.10) are directly proportional to their masses of the solutions. This indicates that the existence of the masses imply the existence of their surface gravities and the temperatures on the horizons. The existence of horizons is also in accord with the cosmological horizon of de Sitter space with constant $\Lambda$ [8], which is considered to be a common candidate of dark energy with the parameter $w= - 1$ [13, 19-26]. This parameter of equation of state is also true for both the cosmological constant $\Lambda$ as well as the cosmological function $\Lambda(u)$ of the {\it rotating} and {\it non-rotating} de Sitter solutions [3]. According to the length scale $r>10^{60}$ suggested by Bousso [13], we find the approximate sizes of the masses less than $(1/2)\times 10^{-60}$, which are bigger than the size of the cosmological constant $|\Lambda|\leq3\times10^{-120}$ with the horizon $r_\Lambda=\sqrt{3/\Lambda}$. It is noted that to the best of the authors knowledge, the solutions are not been seen discussed before. We hope that the exact solutions (2.7) and (3.2) may provide examples of space-times admitting dark energy-momentum tensors (stationary and non-stationary) having negative pressures with the equation of state parameters $w=-1/2$ in the accelerated expanding space-time geometries.

\section*{Acknowledgement}

The authors, Ibohal and Ishwarchandra acknowledge their appreciation for hospitality received from Inter-University Centre for Astronomy and Astrophysics (IUCAA), Pune during their visit in preparing the paper. The work of Ibohal is supported by the University Grants Commission (UGC), New Delhi, File No. 31-87/2005 (SR).

\begin{appendix}
\setcounter{equation}{0}
\renewcommand{\theequation}{A\arabic{equation}}

\section*{Appendix A: Energy conservation equations}

In this appendix we shall show the fact that the energy-momentum
tensor (3.4) satisfies the energy conservation equations
$T^{ab}_{\;\;\;\;;b}=0$. These are four equations, which, using Newman-Penrose (NP) complex spin coefficients [4], can
equivalently be expressed in three equations -- two real and
one complex. Hence, we
find the following
\begin{eqnarray}
&&D\rho=(\rho+ p)(\rho^*+\bar{\rho}^*),\\
&&\delta p=\mu\kappa^*+(\rho+p)(\tau-\bar{\pi}), \\
&&D\mu+\nabla \rho
= \mu\{(\rho^*+\bar{\rho}^*)- 2 (\epsilon+\bar{\epsilon})\}
-(\rho+ p)(\mu^*+\bar{\mu}^*),
\end{eqnarray}
where $\kappa^*$, $\rho^*$, $\mu^*$, $\tau$, $\pi$, etc. are spin coefficients, and
$D, \nabla$ and $\delta$ are the intrinsic derivative operators.
These (A1-A3) are general equations for an energy-momentum tensor of the type (3.4).

Now, in order to verify the conservation equations (A1-A3) for the components of $T^{ab}$ with the quantities $\mu$, $\rho$, $p$ given in (3.3) we present the NP spin coefficients for the  non-stationary metric (3.2):
\begin{eqnarray}
&&\kappa^*=\sigma=\lambda=\epsilon=\pi=\tau=\nu=0, \cr\cr
&&\rho^*=-\frac{1}{r}, \quad
\mu^*=-\frac{1}{2r}\{1-2rm(u)\}, \\
&&\beta=-\alpha={1\over {2\surd 2r}}\,\cot\theta, \quad \gamma=\frac{1}{2}\,m(u). \nonumber
\end{eqnarray}
The intrinsic derivative operators are given as follows:
\begin{eqnarray}
&&D\equiv \ell^a \partial_a=\partial_r, \cr\cr &&\nabla\equiv n^a \partial_a,=\partial_u
-\frac{1}{2}\{1-2rm(u)\}\partial_r , \\
&&\delta \equiv m^a \partial_a=\frac{1}{\surd 2\,r}\Big\{\partial_{\theta} +
\frac{i}{\sin\theta}\,\partial_\phi\Big\}.\nonumber
\end{eqnarray}
where $\ell_a$,\, $n_a$
and $m_a$ are the tetrad null vectors.
The equations (A1)
and (A2) are satisfied by using (A4) and (A5). By virtue of (3.3), (A4) and (A5), we find the left side of (A3) as
\begin{eqnarray}
D\mu+\nabla \rho
=\frac{4}{Kr}\,m(u)_{,u} -\frac{2}{Kr^2}\,m(u)\{1-2rm(u)\}.
\end{eqnarray}
This can be shown equal to the right side of (A3) after using (3.3) and (A4). It leads to the conclusion of the verification that the energy-momentum tensor (3.4) satisfies the energy conservation equations $T^{ab}_{\;\;\;\,;b}=0$. This indicates that the non-stationary line element (3.2) is an exact solution of Einstein's equations. It is also to mention that when $m(u)$ sets to a constant $m$ for the stationary solution (2.7), the energy-momentum tensor (2.8) satisfies the energy conservation equations (2.12).
\end{appendix}

\end{document}